\documentstyle[epsf,psfig,gcdv]{article}

\def\reference{\parskip 0pt\par\noindent\hangindent 0.5 truecm}

\def\spose#1{\hbox to 0pt{#1\hss}}
\def\simlt{\mathrel{\spose{\lower 3pt\hbox{$\mathchar"218$}}
     \raise 2.0pt\hbox{$\mathchar"13C$}}}
\def\simgt{\mathrel{\spose{\lower 3pt\hbox{$\mathchar"218$}}
     \raise 2.0pt\hbox{$\mathchar"13E$}}}

\begin{document}

\small
\shorttitle{Photometric Properties of White Dwarf Dominated Halos}
\shortauthor{H.-c.~Lee et~al.}
\title{\large \bf
Photometric Properties of White Dwarf Dominated Halos
}
\author{\small 
 Hyun-chul Lee$^1$, Brad K. Gibson$^1$, Yeshe Fenner$^1$, Chris B. Brook$^1$,\\
 {\small Daisuke Kawata$^1$, Agostino Renda$^1$, Janne Holopainen$^{1,2}$ \& 
 Chris Flynn$^{1,2}$}
}

\date{}
\twocolumn[
\maketitle
\vspace{-20pt}
\small
{\center
$^1$Centre for Astrophysics \& Supercomputing, 
Swinburne University, Mail \#31, Hawthorn, VIC 3122, Australia\\
hclee,bgibson,yfenner,cbrook,dkawata,arenda,jholopai,cflynn@astro.swin.edu.au\\[1mm]
$^2$Tuorla Observatory, Piikki\"o, FIN-21500, Finland\\
jaolho,cflynn@astro.utu.fi\\[3mm]
}

\begin{center}
{\bfseries Abstract}
\end{center}
\begin{quotation}
\begin{small}
\vspace{-5pt}
Using stellar population synthesis techniques, we explore the
photometric signatures of white dwarf progenitor dominated galactic
halos, in order to constrain the fraction of halo mass that may be
locked-up in white dwarf stellar remnants. We first construct
a 10$^9$~M$_\odot$ stellar halo using the canonical Salpeter 
initial stellar
mass distribution, and then allow for an additional component of low- and
intermediate-mass stars, which ultimately give rise to white dwarf remnants.
Microlensing observations towards the Large Magellanic Cloud, coupled with
several ground-based proper motion surveys, have led to claims that in
excess of 20\% of the dynamical mass of the halo (10$^{12}$~M$_\odot$) 
might be found in white dwarfs.  Our results indicate that (1) even
if only 1\% of the dynamical 
mass of the dark halo today could be attributed to white dwarfs, their
main sequence progenitors at high redshift ($z$$\approx$3) would have
resulted in halos more than 100 times more luminous than those expected 
from conventional initial mass functions alone, and (2) any
putative halo white dwarf progenitor dominated initial mass function component,
regardless of its dynamical importance, would be virtually impossible to 
detect at the present-day, due to its extremely faint surface brightness.
\\
{\bf Keywords:  galaxies: halo ---
galaxies: luminosity function, mass function ---
galaxies: stellar content ---
stars: white dwarfs
}
\end{small}
\end{quotation}
]

\bigskip

\section{Introduction}

The flat rotation curves of spiral galaxies suggest
that their galactic halos are mostly composed of dark matter (Rubin,
Ford \& Thonnard 1980). The precise nature of this dark matter
remains an open question, one with
important implications for cosmology and galaxy formation.
Observations of microlensing events from programs such as MACHO and
EROS suggest that {\it perhaps} of order 20\% of the Galactic halo may be
composed of $\sim$0.5~M$_\odot$ compact objects (Alcock et~al.
2000; Afonso et~al. 2003).  One particularly appealing 
source of 0.5~M$_\odot$ objects that was immediately vetted in the
literature was that of a population of faint, old,
white dwarfs (WDs).\footnote{``Appealing'' in the sense that they are
relatively ``mundane'' astronomical objects, obviating the need to 
draw upon more exotic solutions.}

Alcock et~al. (2000) analysed 5.7 years of photometric data on stars
in the Large Magellanic Cloud, in search of gravitational microlensing
events. They concluded that the fraction of massive compact halo
objects (MACHOs) is about 20\% (with an average mass of 
0.5~M$_\odot$ per MACHO), and that the total mass in MACHOs out to 50~kpc is
10$^{11}$~M$_\odot$, assuming the lenses are
located in the Galactic halo.\footnote{Not necessarily a unanimously
accepted hypothesis - e.g. Sahu \& Sahu (1998); Di~Stefano (2000).}
The recent EROS data
toward the Small Magellanic Cloud allow for a maximum of
$\sim$25\% of a spherical, isothermal, and isotropic Galactic halo
of mass 4$\times$10$^{11}$~M$_\odot$ out to 50~kpc which could
be composed of objects with mass between 2$\times$10$^{-7}$~M$_\odot$ 
and 1~M$_\odot$ (Afonso et~al. 2003). The absence of events with
crossing times shorter than 10~days and the lack of sufficient numbers of
low-mass main sequence stars in the Hubble Deep Field 
essentially rules out
planet-like objects and brown dwarfs as the microlens candidates (Gould,
Bahcall \& Flynn 1998; Lasserre et~al. 2000).  The current favoured
mass range from the microlensing experiments also rules out the
remnants of more massive stars, including
neutron stars and black holes, as potential (and substantial)
baryonic dark matter candidates.

A rich literature has emerged over the past five years exploring the
pros and cons of the hypothesis that white dwarfs may comprise a 
significant fraction of the dynamical mass of the galaxies.
Ryu, Olive \& Silk (1990) constructed a simple
galactic halo chemical evolution model and ruled out neutron stars,
but not white dwarfs, as dark matter candidates, based on metallicity
and luminosity considerations. The low incidence of Type~Ia supernovae
observed in the outer regions of galaxies lead Smecker \& Wyse (1991)
to impose tight upper limits on the fraction of white dwarfs in
galactic halos.  Charlot \& Silk (1995), using population synthesis
models coupled with number counts of faint galaxies in deep galaxy
surveys, showed that only a small fraction ($\simlt$10\%) of
present-day halos could be in the form of white dwarfs.
Madau \& Pozzetti (2000) showed that in order to avoid the
overproduction of extragalactic background light, the halo white dwarf
mass fraction should be $\simlt$5\%.  Gibson \& Mould (1997) and Brook,
Kawata \& Gibson (2003) placed more severe 
constraints upon the halo white dwarf mass fraction 
($\simlt$1--2\%) by using the observed carbon, nitrogen, and oxygen 
abundance patterns of halo stars.

Following Larson's (1986) suggestion that a remnant-dominated form of
the initial mass function (IMF) could account for the unseen mass in the
solar neighborhood, both Chabrier, Segretain \& Mera (1996) and Adams \&
Laughlin (1996) devised physically-motivated models allowing for a
white dwarf progenitor dominated IMF, consistent with the aforementioned
MACHO and EROS microlensing results.  Such an IMF differs from that of the
classical Salpeter (1955) functional form through the absence of both
low mass ($\simlt$1~M$_\odot$) and high mass ($\simgt$6~M$_\odot$) stars
(Gibson \& Mould 1997; Fig~1).

In this paper, we address explicitly the photometric properties of the
luminous early phases of putative white dwarf progenitor dominated galactic
halos, in addition to their temporal evolution. In Section~2, we describe our
stellar population models and the different functional
forms for the IMFs considered here.
The results of our calculations are then presented in
Section~3, with the accompanying discussion and conclusions provided
in Section~4.

\section{Models}

Using our evolutionary stellar population synthesis code
(Lee, Lee \& Gibson 2002), we have
calculated the evolution of the photometric properties of galactic
halos as a function of the dynamical mass fraction tied up in putative
populations of white dwarfs.  We allow the total
present-day mass fraction of white dwarfs to range from 0 -- 100\%
\footnote{Strictly speaking, we allow up to 99.9\% present-day mass fraction 
of white dwarfs as a 10$^9$~M$_\odot$ stellar halo with the 
canonical Salpeter IMF provides the base for our model.} of
the halo's dynamical mass, which is taken for this exercise 
to be 10$^{12}$~M$_\odot$ (after that of the Milky Way - 
Fich \& Tremaine 1991). All models described herein 
contain a standard 10$^{9}$~M$_\odot$ stellar component\footnote{
A stellar mass of 10$^{9}$~M$_\odot$ for the Milky Way halo is reasonable once
one adopts, for example, a surface density profile proportional to r$^{-2}$
normalized by the local stellar halo density - e.g. Preston et~al.
(1991).}, which is
itself described by the Salpeter IMF (by number):
\begin{equation}
\Phi (m){\rm d}m = dn/dm = Am^{-x}\,{\rm d}m,
\end{equation}
with $x$=2.35.

The fraction of halo dark matter contained in white dwarfs is then
determined by the amount of matter contained in a supplemental IMF
described by a truncated power-law of the form
\begin{equation}
\Phi (m)\,{\rm d}m = dn/dm = Ae^{-\left({\bar{m}/m}\right)^{\beta}}\times
m^{-\alpha}\,{\rm d}m
\end{equation}
\noindent
for which we use $\bar{m}=2.7$, $\beta=2.2$, and $\alpha=5.75$
(Chabrier et~al. 1996).  The
peak of this skewed-Gaussian functional form for the IMF 
(hereafter, wdIMF) occurs at 
$m\approx$2~M$_\odot$, favouring the production of white dwarf progenitors.
The wdIMF yields a present day Galactic halo
mass-to-light ratio $\gg$ 100 after a Hubble time as most of its initial
stellar distribution has since become very faint remnants.

The stellar population synthesis models presented here are based upon
the $Y^{2}$ Isochrones\footnote{\tt
http://csaweb.yonsei.ac.kr/$\sim$kim/yyiso.html\rm} (Kim et~al. 2002)
with [$\alpha$/Fe]=$+$0.3, coupled to the post-red giant branch
stellar evolutionary tracks of Yi, Demarque \& Kim (1997). We have
calibrated the horizontal-branch morphology with the Milky Way
globular clusters, as in Lee et~al. (2000, 2002).  The stellar library
of Lejeune, Cuisinier \& Buser (1998) was taken for the conversion
from theoretical to observable quantities.
In our calculations, the stellar remnant mass was assumed to be 
0.5~M$_\odot$ for initial stellar masses below 8~M$_\odot$, and
1.0~M$_\odot$ for initial stellar masses in excess of 8~M$_\odot$.

New cooling models for white dwarfs have become available recently which
possess somewhat different behaviour in the colour-magnitude plane at
advanced ages (e.g. Hansen 1998; Richer et~al. 2000),
when compared with the conventional models
adopted in most population sysnthesis codes (including ours).
While we have experimented with their inclusion, due to the extremely
low luminosity at which these new models diverge from the classical ones
(at $M_{V}$$\sim$17.5, after $\sim$8~Gyr of cooling), our results are
not seriously impacted by the choice of specific white dwarf cooling
tracks.

The following section presents results for our default model -
a Milky Way-like halo with 
metallicity [Fe/H] $\sim$ $-$1.8 (Ryan \& Norris 1991) - we also
touch briefly
upon the implications for more metal-rich halos such as that for M31.

\section{Results}

Figure~1 shows the variation in absolute V-band magnitude as a
function of halo dynamical mass fraction tied up in
white dwarfs, for ages ranging from 1 to
14~Gyr.  The asymptotic values of M$_{\rm V}$
approached for mass fractions $<$10$^{-4}$ simply reflect 
the model's assumed base 10$^9$~M$_\odot$ Salpeter IMF stellar
component;\footnote{We re-emphasie here that the asymptotic value of 
M$_{\rm V}$ approached at low white dwarf mass fractions in Figure~1
corresponds to the values expected for a halo with a base stellar component of
mass 10$^9$~M$_\odot$, constructed with a conventional Salpeter IMF -
representative of what one might expect for a canonical model of 
the Milky Way.  This absolute magnitude would of course change should a 
different ``base'' Salpeter IMF halo component be employed - for example
M$_{V}$ for a halo of the same metallicity 
would be 2.5~mag brighter should the base stellar halo component have a mass
of 10$^{10}$~M$_\odot$.
Furthermore, the halo would be about 1~mag fainter if its metallicity
was solar, as the main-sequence turnoff is fainter
for metal-rich systems.  Regardless, our analysis remains valid, as it
is the \it differential \rm between the models which is important, and \it not
\rm the ``zero-point'' of the curves in Figure~1.}
a mass fraction of unity corresponds to a 10$^{12}$~M$_\odot$ 
halo comprised entirely (essentially) of white dwarfs.  After $\sim$13~Gyr, a
100\% white dwarf halo is, for all intents and purposes,
indistinguishable photometrically from a halo containing no 
remnants aside from the underlying population attributed to the (known)
Salpeter-like component.  For younger stellar systems though, 
WD progenitors preferentially populate the main
sequence, giant branch, and horizontal branch phases and rapidly
dominate the luminosity of the halo.

\begin{figure}
\begin{center}
  \leavevmode
  \epsfxsize 8cm
  \epsfysize 8cm
  \epsffile{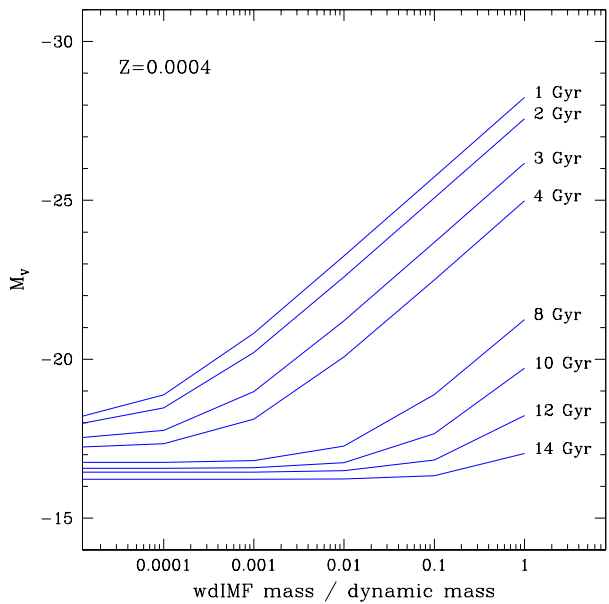}
\end{center}
\caption{
  The absolute V-band magnitude as a function of halo white
  dwarf mass fraction for eight ages ranging from 1 to 14~Gyr. 
  The left-hand limit to the plot corresponds to the minimum
  contribution due to the known 10$^9$~M$_\odot$ stellar halo
  intrinsic to spirals such as the Milky Way (our default model).
  Passive evolution of this basic system would be $\sim$2 magnitudes
  brighter at redshift $z$$\sim$3 (an age of $\sim$1~Gyr, for
  a redshift of formation of five).
  Conversely, a 10$^{12}$~M$_\odot$ halo comprised of nothing
  but white dwarfs at the present-day would have been $\sim$10
  magnitudes brighter at $z$$\sim$3.
}
\end{figure}

Although unlikely, if the total dynamic mass were composed of white
dwarf remnants, this halo would have been $\sim$10 magnitudes brighter 
$\sim$1~Gyr after its ``formation'' (i.e. $z$$\sim$3, for a redshift
of formation of five).  In comparison, the same halo (by dynamical mass),
but now with only the known 10$^9$~M$_\odot$ Salpeter 
IMF stellar component, would only be $\sim$2 magnitudes brighter at
the same lookback time. Even if only 1\% of the
mass of the dark halo was in the form of this wdIMF component, 
it would have been a factor of $\sim$100 times more luminous than
the canonical halo at the same redshift $z$$\sim$3.

\begin{figure}
\begin{center}
  \leavevmode
  \epsfxsize 8cm
  \epsfysize 8cm
  \epsffile{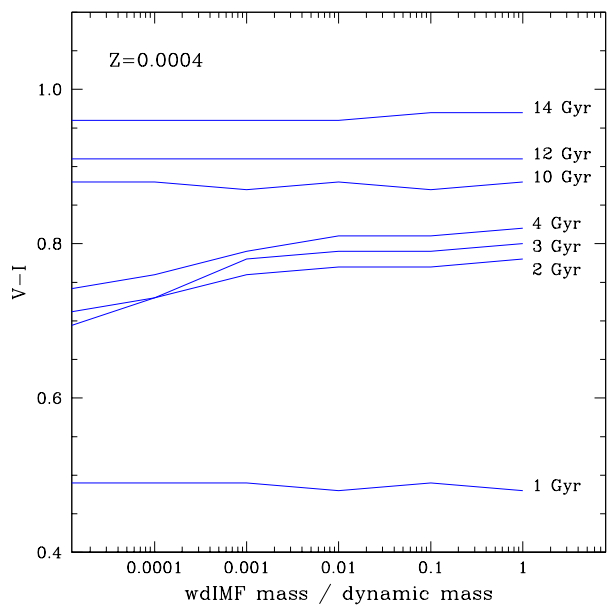}
\end{center}
\caption{
  As in Figure~1, but now showing the variation in integrated
  colour (V$-$I) as a function of halo white dwarf mass fraction.
}
\end{figure}

Figure~2 presents the dependence of colour (V$-$I) on the fraction of
the dynamic halo mass in the form WD remnants for halos ranging in
age from 1 to 14~Gyr.  A white dwarf dominated halo is indistinguishable
in colour from a Salpeter-only halo for lookback times $\simlt$8~Gyr
(note the essentially flat behaviour of the curves as a function of
wdIMF mass fraction).\footnote{The abrupt shift to significantly
bluer colours for the 1~Gyr isochrone
is due to the treatment of convective core overshooting in
the stellar models (Yi 2003; Fig~6).}

The only marginal difference between the
two scenarios occurs at ages corresponding to redshifts in the range
1$\simlt$$z$$\simlt$2 - for
instance, if 1\% of the dark halo mass was in the form of WDs, its stellar
component would
have been $\sim$0.08 mag $\it{redder}$ than a Salpeter-only
halo (left-hand limit of Figure~2) at early epochs
because of the former's more significant populations
of red giants and red horizontal-branch stars.

The predicted colours in Figure~2 are also, of course, different for
different metallicities. In general, the more metal-rich a halo is, the
redder its stellar population. For a halo with a
metallicity like that of M31 ([Fe/H] $\sim$ $-$0.5 - Durrell, 
Harris \& Pritchet 2001), we note in passing that 
if 1\% of the halo's dynamical mass were attributed to a
wdIMF component, the V$-$I colour at 1~Gyr would be 0.55, at 3~Gyr it
would be 1.14, and at 14~Gyr it would be 1.26.

\section{Discussion}

In Figure~3 (after Chabrier 2001), 
we speculate on the surface brightness radial profile of
a 10$^{12}$~M$\odot$ halo with a 10\% (by mass) wdIMF component, but
\emph{without} the ``base'' 10$^{9}$~M$_\odot$ Salpeter IMF stellar 
component.\footnote{By eliminating the underlying trace (by mass, not
by luminosity) Salpeter component, we serve to emphasise the light contribution
from the wdIMF component {\it alone}.}
We have adopted a dark matter halo density profile of
$\rho(r) \propto 1/r^2$.
For the surface mass density of the halo, we 
assumed 23~M$_\odot/pc^{2}$ at 8~kpc (Kuijken \& Gilmore 1991). It can be seen
from Figure~3 that if 10\% of the dark halo mass were composed of a
wdIMF stellar component, its predicted surface brightness would be 
essentially undetectable.  Put into context, the central
surface brightness of the low surface brightness galaxy Malin~1 
is $\mu_\circ$(V)$\sim$25.5 (Bothun et~al.
1987) - i.e., this 10\% wdIMF component lies 7--10 magnitudes {\it
fainter} than the surface brightness of Malin~1!

\begin{table}[htp]
\caption{Surface Brightness at 60~kpc (Z=0.0004) }
\begin{center}
\begin{tabular}{ccccc}
  \hline
  Age (wdIMF) & B & V & R & I \\ 
  \hline
  12~Gyr (1\%) & 38.70 & 37.44 & 36.59 & 35.78 \\
  12~Gyr (10\%) & 36.20 & 34.94 & 34.09 & 33.28 \\
  12~Gyr (100\%) & 33.70 & 32.44 & 31.59 & 30.78 \\
  14~Gyr (1\%) & 40.45 & 39.12 & 38.22 & 37.38 \\
  14~Gyr (10\%) & 37.95 & 36.62 & 35.72 & 34.88 \\
  14~Gyr (100\%) & 35.45 & 34.12 & 33.22 & 32.38 \\
  \hline
 \end{tabular}
\end{center}
\end{table}

\begin{figure}
\begin{center}
  \leavevmode
  \epsfxsize 8cm
  \epsfysize 8cm
  \epsffile{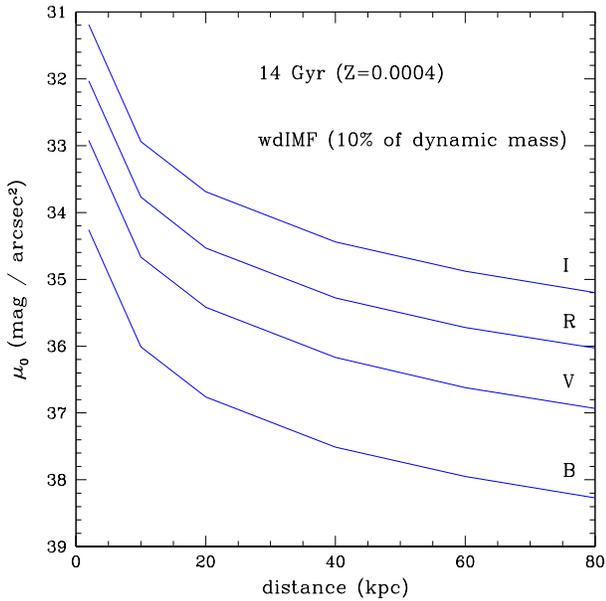}
\end{center}
\caption{
Surface brightness radial profile of a galactic halo with 10\% of its
dynamical mass comprised of a wdIMF stellar component, but now
\emph{without} the base 10$^{9}$~M$_\odot$ Salpeter IMF stellar component
- to emphasise the essentially undetectable wdIMF component's
surface brightness - after Chabrier (2001).
}
\end{figure}

Table~1 lists the surface brightness of such a galactic halo at 
a galactocentric radius of 60~kpc
for several passbands at a given age and wdIMF mass fraction.
We do not wish to belabour the result here, and simply note that the
surface brightnesses in the outer regions of these putative WD-dominated
halos would be essentially undetectable.
More metal-rich halos would be somewhat
brighter because the main-sequence lifetime of stars increases with
stellar metallicity, but still well below any realistically observable limit.

\section{Summary}

We have explored the photometric ramifications of ascribing a significant
fraction of a galactic halos dynamical mass to a population of old, faint,
white dwarfs.  Using our population synthesis package, we show that
while such an hypothesis is essentially impossible to rule out using
surface brightness arguments for nearby (old) halos, passive evolution
of the progenitors of any large, putative, halo white dwarf population
would make the present-day halos 100 -- 500 times brighter at 
redshifts 1$\simlt z \simlt$3 than halos evolving passively under
standard stellar evolution arguments (i.e. a Salpeter-like initial mass
function).  

Making one-to-one comparisons between
single objects such as the Milky Way and potentially quite different
objects at high-redshift can be dangerous; what we wish to end with
is simply a note drawing attention to, for example, the existence of
passively-evolving L$^{*}$ galaxies at high-redshift which \it individually \rm
are consistent with passive evolution of normal stellar populations
and normal IMFs (e.g. Waddington et al. 2002) -
i.e. the stellar populations therein are $<$ 2 mags brighter then
than today (under reasonable assumptions of mass, etc), opposed to
being $>$ 4-5 mags brighter than expected if these galaxies were
dominated dynamically by white dwarf precursors.  A safer comparison
is to model the ensemble of the population (e.g. extragalactic
background light and deep number counts), both of which are also
inconsistent with the white dwarf scenario (Charlot \& Silk 1995;
Madau \& Pozzetti 2000).

\section*{Acknowledgements}

We wish to acknowledge the Australian Research Council,
through its Discovery Project and Linkage International schemes,
for ongoing financial support.  We are grateful to the anonymous referees 
for their detailed reports that helped us improve this paper.
We also thank Gilles Chabrier for stimulating our pursuit of this result.

\section*{References}

\reference Adams, F.C. \& Laughlin, G. 1996, ApJ, 468, 586
\reference Afonso, C., Albert, J.N., Andersen, J., et~al. 2003, A\&A, 400, 951
\reference Alcock, C., Allsman, R.A., Alves, D.R., et~al. 2000, ApJ, 542, 281
\reference Bothun, G.D., Impey, C.D., Malin, D.F. \& Mould, J.R. 1987,
AJ, 94, 23
\reference Brook, C., Kawata, D. \& Gibson, B.K. 2003, MNRAS, 343, 913
\reference Chabrier, G. 2001, in Richer, H.B. \& Gibson, B.K., eds.,
White Dwarfs as Dark Matter,\hfill\break
{\tt http://www.astro.ubc.ca/WD\_workshop/talks/}
\reference Chabrier, G., Segretain, L. \& Mera, D. 1996, ApJ, 468, L21
\reference Charlot, S. \& Silk, J. 1995, ApJ, 445, 124
\reference Di~Stefano, R. 2000, ApJ, 541, 587
\reference Durrell, P.R., Harris, W.E. \& Pritchet, C.J. 2001, AJ, 121, 2557
\reference Fich, M. \& Tremaine, S. 1991, ARAA, 29, 409
\reference Gibson, B.K. \& Mould, J.R. 1997, ApJ, 482, 98
\reference Gould, A., Flynn, C. \& Bahcall, J.N. 1998, ApJ, 503, 798
\reference Hansen, B.M.S. 1998, Nature, 394, 860
\reference Kim, Y.-C., Demarque, P., Yi, S.K. \& Alexander, D.R. 2002,
   ApJS, 143, 499
\reference Kuijken, K. \& Gilmore, G. 1991, ApJ, 367, L9
\reference Larson, R.B. 1986, MNRAS, 218, 409
\reference Lasserre, T., Afonso, C., Albert, J.N., et~al. 2000, A\&A, 355, L39
\reference Lee, H.-c., Lee, Y.-W. \& Gibson, B.K. 2002, AJ, 124, 2664
\reference Lee, H.-c., Yoon, S.-J. \& Lee, Y.-W. 2000, AJ, 120, 998
\reference Lejeune, T., Cuisinier, F. \& Buser, R. 1998, A\&AS, 130, 65
\reference Madau, P. \& Pozzetti, L. 2000, MNRAS, 312, L9
\reference Preston, G.W., Shectman, S.A. \& Beers, T.C. 1991, ApJ, 375, 121
\reference Ryan, S.G. \& Norris, J.E. 1991, AJ, 101, 1865
\reference Richer, H.B., Hansen, B., Limongi, M., et~al., 2000, ApJ, 529, 318
\reference Rubin, V.C., Ford, W.K.Jr. \& Thonnard, N. 1980, ApJ, 238, 471
\reference Ryu, D., Olive, K.A. \& Silk, J. 1990, ApJ, 353, 81
\reference Sahu, K.C. \& Sahu, M.S. 1998, ApJ, 508, L147
\reference Salpeter E.E. 1955, ApJ, 121, 161
\reference Smecker, T.A. \& Wyse, R. 1991, ApJ, 372, 448
\reference Waddington, I., Windhorst, R.A., Cohen, S.H., et~al. 2002, 
   MNRAS, 336, 1342
\reference Yi, S. 2003, ApJ, 582, 202
\reference Yi, S., Demarque, P. \& Kim, Y.-C. 1997, ApJ, 482, 677

\end{document}